\documentclass{aa} 
\usepackage[varg]{txfonts}
\usepackage{graphicx}
\usepackage{natbib}
\newcommand{\isotope}[2]{${}^{#1}$#2}

\newcommand{\msunb}{\mbox{$\mathrm{M_{\odot}}$} }

\newcommand{\napo}{$^{16}$O$($p,$\alpha)^{13}$N }
\newcommand{\photo}{$^{16}$O$(\gamma,\alpha)^{12}$C }
\newcommand{\chain}{$^{16}$O$($p,$\alpha)^{13}$N$(\gamma$,p$)^{12}$C }
\newcommand{\ofus}{$^{16}\mathrm{O}+^{16}\mathrm{O}$ }

\newcommand{\cofus}{$^{12}\mathrm{C}+^{16}\mathrm{O}$ }
\newcommand{\mnapo}{\mathrm{Op}\alpha}
\newcommand{\mphoto}{\mathrm{O}\gamma\alpha}
\newcommand{\mofus}{\mathrm{O+O}}
\newcommand{\mcfus}{\mathrm{C+C}}
\newcommand{\mcofus}{\mathrm{C+O}}

\begin{document}
   \title{\napo makes explosive oxygen burning sensitive to the metallicity of the progenitors of type Ia supernovae}
   \titlerunning{\napo boosts $\alpha$-rich explosive O-burning}

   \author{E. Bravo
   }

   \institute{E.T.S. Arquitectura del Vall\`es, Universitat Polit\`ecnica de Catalunya, Carrer Pere Serra  
1-15, 08173 Sant Cugat del Vall\`es, Spain\\   
              \email{eduardo.bravo@upc.edu} \label{inst1}}

   \date{Received ; accepted }

 
  \abstract{
Even though the main nucleosynthetic products of type Ia supernovae belong to the iron-group, intermediate-mass alpha-nuclei (silicon, sulfur, argon, and calcium) stand out in their spectra up to several weeks past maximum brightness. Recent measurements of the abundances of calcium, argon, and sulfur in type Ia supernova remnants have been interpreted in terms of metallicity-dependent oxygen burning, in accordance with previous theoretical predictions. It is known that $\alpha$-rich oxygen burning results from \isotope{16}{O}$\rightarrow$\isotope{12}{C} followed by efficient \isotope{12}{C}+\isotope{12}{C} fusion reaction, as compared to oxygen consumption by \isotope{16}{O} fusion reactions, but the precise mechanism of dependence on the progenitor metallicity has remained unidentified so far. I show that the chain $^{16}$O(p,$\alpha$)$^{13}$N($\gamma$,p)$^{12}$C boosts $\alpha$-rich oxygen burning when the proton abundance is large, increasing the synthesis of argon and calcium with respect to sulfur and silicon. For high-metallicity progenitors, the presence of free neutrons leads to a drop in the proton abundance and the above chain is not efficient. Although the rate of \napo can be found in astrophysical reaction rate libraries, its uncertainty is unconstrained. Assuming that all reaction rates other than \napo retain their standard values, an increase by a factor of approximately seven of the \napo rate at temperatures in the order $3-5\times10^9$~K is enough to explain the whole range of calcium-to-sulfur mass ratios measured in Milky Way and LMC supernova remnants. These same measurements provide a lower limit to the \napo rate in the mentioned temperature range, on the order of a factor of 0.5 with respect to the rate reported in widely used literature tabulations. 
}


\maketitle
%

\section{Introduction}

The nucleosynthesis resulting from type Ia supernovae (SNIa) reflects the thermodynamical history of the progenitor white dwarf (WD) during the explosion and its initial chemical composition. Thus, nucleosynthetic constraints coming from observations of supernovae and their remnants are an important source of knowledge of the conditions achieved during the explosion. The optical properties, spectra, and light curves of SNIa over a few weeks around maximum brightness have been used to infer the chemical profile of the ejecta \citep{2005sth,2008maz,2011tan,2014sas,2016ash}. However, the ability to constrain the nucleosynthetic products based on optical data is hampered by the complex physics that governs the formation of spectral features in the visible, ultraviolet, and infrared bands. 

Observations of sufficiently close supernova remnants (SNRs) are an alternative to obtain information about the chemical composition of the ejecta \citep[e.g.][]{1988ham,1988fes}. Hundreds to a few thousands of years after the explosion, the ejected elements emit strongly in the X-ray band due to shock heating, and their emission lines can be detected and measured by current X-ray observatories \citep[e.g.][]{1995hug,1995van,2008bad,2014yam,2015yam}. Recently, the high spectral resolution of {\it Suzaku} has allowed the relative mass ratio of calcium to sulfur, $M_\mathrm{Ca}/M_\mathrm{S}$, to be measured in a few
SNRs with a precision of $\sim5\%-16\%$ \citep{2017mar}, with the result that this ratio spans the range $0.17 - 0.28$, with an uncertainty of 0.04 in both limits \citep[for reference, this mass ratio is 0.177 in the solar system;][]{2003lod}. These results have been interpreted in terms of metallicity-dependent yields during explosive oxygen burning.

There are two effects to account for in relation with $\alpha$-rich oxygen burning: first, the strength of the enhancement of the yield of calcium at all metallicities, and second, the metallicity dependence of the mass ratio of calcium to sulfur, $M_\mathrm{Ca}/M_\mathrm{S}$, in the ejecta.
Both calcium and sulfur are a product of explosive oxygen burning, and they are synthesized in proportion to their ratio in conditions of quasi-statistical equilibrium, which depends on the quantity of $\alpha$ particles available:  $M_\mathrm{Ca}/M_\mathrm{S}\propto X_\alpha^2$ \citep{2014de}. \cite{1973woo} studied the conditions under which explosive oxygen burning would reproduce the solar-system abundances. They explained that oxygen burning can proceed through two different branches: $\alpha$-poor and $\alpha$-rich. The $\alpha$-poor branch has the net effect that for every two \isotope{16}{O} nuclei destroyed,  one \isotope{28}{Si} nuclei and one $\alpha$ particle are created. This branch proceeds mainly through the fusion reaction of two \isotope{16}{O} nuclei, but it is contributed as well by the chain \isotope{16}{O}$(\gamma,\alpha)$\isotope{12}{C}$($\isotope{16}{O}$,\gamma)$\isotope{28}{Si}. On the other hand, the $\alpha$-rich branch involves the photo-disintegration of two \isotope{16}{O} nuclei to give two \isotope{12}{C} plus two $\alpha$ particles, followed by the fusion reaction \isotope{12}{C}$($\isotope{12}{C}$,\alpha)$\isotope{20}{Ne}$(\gamma,\alpha)$\isotope{16}{O}, which releases a total of four $\alpha$ particles for each \isotope{16}{O} nuclei destroyed. \cite{1973woo} included the chain \chain in the $\alpha$-rich branch and listed these two reactions (and their inverses) among the most influential reactions for explosive oxygen burning. \cite{2012bra} found that the \napo reaction rate and its inverse are among the ones that impact most the abundance of \isotope{40}{Ca}, in agreement with \cite{1973woo}.

\citet{2014de} and \citet{2016mil} noticed that $M_\mathrm{Ca}/M_\mathrm{S}$ can be used to infer the metallicity, $Z$, of the progenitor of SNIa, but they did not identify the source of the metallicity dependence of the calcium and sulfur yields. Later, \citet{2017mar} used the measured $M_\mathrm{Ca}/M_\mathrm{S}$ in a few type Ia SNRs of the Milky Way and the LMC to determine the progenitor metallicity, and concluded that there had to be an unknown source of neutronization of the WD matter before the thermal runaway besides that produced during carbon simmering \citep{2008chm,2008pir,2016mar,2017pie}. They also pointed out that SNIa models that used the standard set of reaction rates were unable to reproduce the high calcium-to-sulfur mass ratio measured in some remnants.

In the present work, it is shown that the origin of the metallicity dependence of $M_\mathrm{Ca}/M_\mathrm{S}$ has to be ascribed to the \napo reaction. In the following section,  the mechanisms by which the \napo reaction controls the $\alpha$ particle abundance as a function of the progenitor metallicity  are explained. If the \napo reaction is switched off, the value of $M_\mathrm{Ca}/M_\mathrm{S}$ remains insensitive to metallicity. In Section~\ref{s:limits}, the uncertainty of the \napo rate is reported along with the limits to its value that can be obtained from the measured $M_\mathrm{Ca}/M_\mathrm{S}$ in SNRs. The conclusions of this work are presented in Section~\ref{s:conclusions}.

\section{\napo and metallicity}\label{s:workings}

The chain $^{16}$O(p,$\alpha$)$^{13}$N($\gamma$,p)$^{12}$C provides a route alternative to \photo to convert \isotope{16}{O} to \isotope{12}{C} and feed the $\alpha$-rich branch of explosive oxygen burning \citep{1973woo}. The chain neither consumes nor produces protons, however its rate depends on the abundance of free protons. In the shells that experience explosive oxygen burning in SNIa, the neutron excess is closely linked to the progenitor metallicity. At small neutron excess, hence low progenitor metallicity, there are enough protons to make \napo operational. At large neutron excess, hence large progenitor metallicity, the presence of free neutrons neutralizes the protons, and undermines the chain \chain efficiency. This is because in explosive oxygen burning, quasi-statistical equilibrium holds for the abundances of nuclei between silicon and calcium \citep{1970tru}. In quasi-statistical equilibrium, a large neutron-excess leads to a large abundance of neutronized intermediate-mass nuclei such as, for instance, $^{34}$S or $^{38}$Ar, which react much more efficiently with protons than the $\alpha$-nuclei such as $^{32}$S or $^{36}$Ar that are produced in low-neutron-excess conditions. 

\begin{figure}
\resizebox{\hsize}{!}{\includegraphics{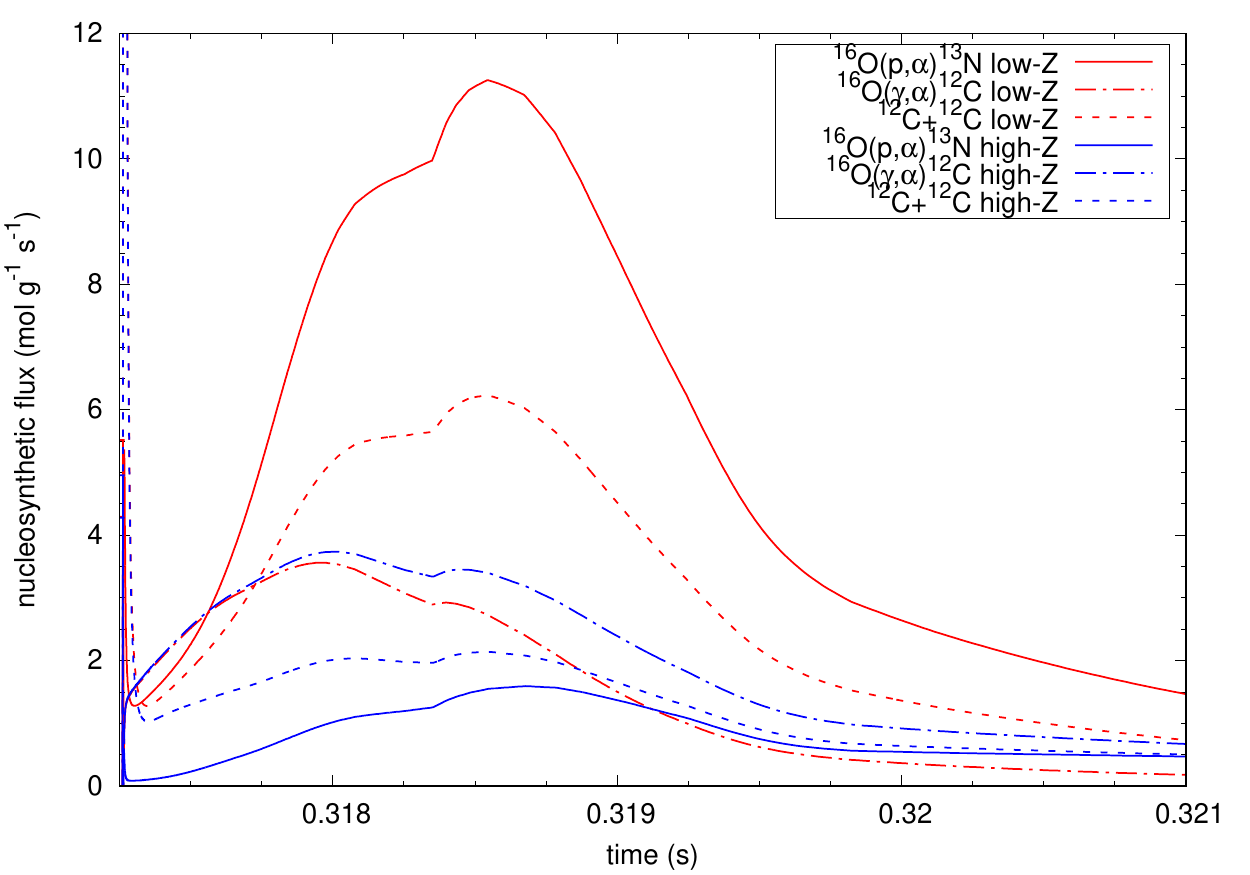}}
\caption{Nucleosynthetic fluxes from \isotope{16}{O} to \isotope{12}{C} due to the \napo (solid lines) and the \photo (dot-dashed lines) reactions as a function of time, in a mass shell with peak temperature $4\times10^9$~K, for a 1.06~\msunb WD with either progenitor metallicity $Z=2.25\times10^{-4}$ (red) or $Z=0.0225$ (blue). The nucleosynthetic flux of the \isotope{12}{C} fusion reaction is also plotted with dotted lines. 
}
\label{f:1}
\end{figure}

\begin{figure*}
\resizebox{\hsize}{!}{\includegraphics{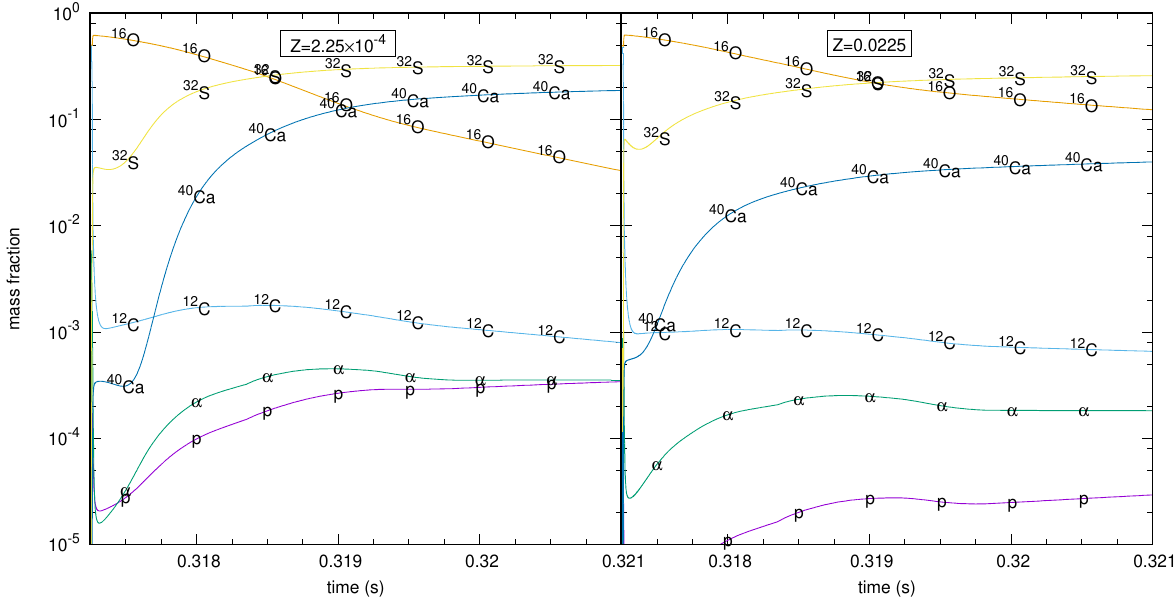}}
\caption{Chemical composition of a detonated mass shell in the same conditions as in Fig.~\ref{f:1}, with progenitor metallicity $Z=2.25\times10^{-4}$ (left) or $Z=0.0225$ (right). The plot shows the initial phases of oxygen burning, starting shortly after the shell was hit by the detonation wave and ending when the oxygen abundance had declined significantly. The peak temperature was reached at time 0.3188~s. 
}
\label{f:2}
\end{figure*}

\begin{figure}
\resizebox{\hsize}{!}{\includegraphics{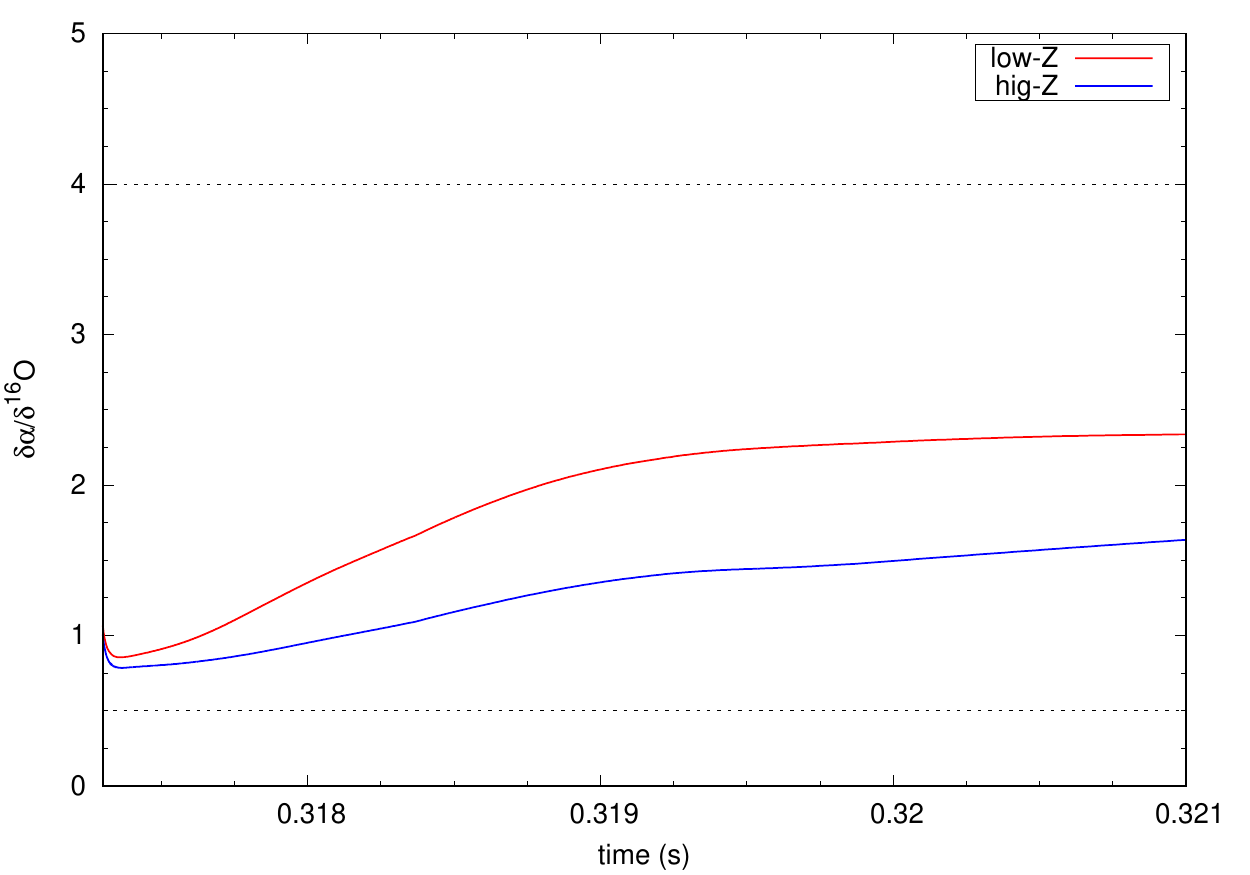}}
\caption{$\alpha$-efficiency of explosive oxygen burning for the same mass shells depicted in Figs.~\ref{f:1} and \ref{f:2}, for the low-$Z$ case (in red) and the high-$Z$ case (in blue). The limit efficiencies for $\alpha$-rich oxygen burning, $\delta\alpha/\delta ^{16}\mathrm{O}=4$, and $\alpha$-poor oxygen burning, $\delta\alpha/\delta ^{16}\mathrm{O}=0.25$, are drawn as dotted lines.
}
\label{f:3}
\end{figure}

To illustrate the above ideas, Figs.~\ref{f:1}-\ref{f:3} show the evolution of key quantities related to the branching of explosive oxygen burning into either the $\alpha$-rich or the $\alpha$-poor tracks. Specifically, the plots show the evolution of a mass shell reaching a peak temperature of $4\times10^9$~K in models 1p06\_Z2p25e-4\_$\xi_\mathrm{CO}$0p9 and 1p06\_Z2p25e-2\_$\xi_\mathrm{CO}$0p9, described in \citet{2019bra}. In short, both models simulate the detonation of a WD with mass $1.06$~\msunb made of carbon and oxygen, whose progenitor metallicities are respectively $Z=2.25\times10^{-4}$ (strongly sub-solar metallicity, hereafter the low-$Z$ case) and $Z=0.0225$ (about 1.6 times solar, hereafter the high-$Z$ case). In both models, the rate of the fusion reaction \cofus has been scaled down by a factor 0.1 as suggested by \citet[][see also Bravo et al. 2019]{2017mar}.

A larger proton abundance in the low-$Z$ case at the same temperature and similar oxygen abundance as in the high-$Z$ case implies a larger nucleosynthetic flux from the \chain chain, as can be seen in Fig.~\ref{f:1}, and as a consequence the nucleosynthetic flux from this reaction chain exceeds that from the \photo reaction. Thus, the \napo reaction becomes the main source of \isotope{12}{C} at the expense of \isotope{16}{O}. In the high-$Z$ case, the nucleosynthetic flux due to the \chain chain remains at all times below that due to the \photo reaction. 

Figure~\ref{f:2} shows the evolution of the abundances of selected nuclei during the main phase of oxygen burning of the aforementioned mass shell, for both metallicities. The low-$Z$ case displays a proton abundance larger than the high-$Z$ case by a factor of approximately ten, while the mass fractions of $\alpha$ particles and \isotope{12}{C} nuclei are also larger by a factor of approximately two.  The abundance of oxygen declines faster in the low-$Z$ case, and that of sulfur rises faster at first, but at the end achieves nearly the same equilibrium abundance as in the high-$Z$ case. In contrast, the mass fraction of calcium rises in the low-$Z$ case to approximately five times the value reached in the high-$Z$ case. The final values of the calcium-to-sulfur mass ratios obtained in the mass shell are: $M_\mathrm{Ca}/M_\mathrm{S}=0.45$ in the low-$Z$ case, and $M_\mathrm{Ca}/M_\mathrm{S}=0.17$ in the high-$Z$ case. 

Figure~\ref{f:3} shows the $\alpha$-efficiency of oxygen burning for both the low-$Z$ and the high-$Z$ case. For the purposes of the present work, the $\alpha$-efficiency is defined as the number of $\alpha$ particles created through both the $\alpha$-rich and the $\alpha$-poor branches divided by the number of \isotope{16}{O} nuclei destroyed in the same processes, and is equal to:
\begin{equation}
 \frac{\delta\alpha}{\delta ^{16}\mathrm{O}} = \frac{R_{\mnapo}+R_{\mphoto}+R_{\mofus}+2R_{\mcfus}}{R_{\mnapo}+R_{\mphoto}+2R_{\mofus}+R_{\mcofus}-R_{\mcfus}}\,,
\end{equation}
\noindent where $R$ is the nucleosynthetic flux due to a given reaction in \mbox{\rm mol~g$^{-1}$~s$^{-1}$},  $R_{\mnapo}=\rho N_\mathrm{Av}\left<\sigma v\right>Y(^{16}\mathrm{O})Y(\mathrm{p})$ makes reference to the \napo reaction and $R_{\mphoto}$ to the \photo photodisintegration, $Y$ is the molar fraction of each species involved in the reaction, $R_{\mofus}$ makes reference to the fusion reaction \ofus, and so on. As explained before, the $\alpha$-efficiency of $\alpha$-poor oxygen burning is 0.5, which would correspond, for instance, to all reaction rates being zero, except that of \ofus. The $\alpha$-efficiency of $\alpha$-rich oxygen burning is equal to 4, which would be obtained if $R_{\mofus} = R_{\mcofus} = 0$, and $R_{\mcfus}=\left(R_{\mnapo}+R_{\mphoto}\right)/2$. In Fig.~\ref{f:1}, it can be seen that $R_{\mcfus}$ is close to the mean of $R_{\mnapo}$ and $R_{\mphoto}$. As could be expected, the $\alpha$-efficiency shown in Fig.~\ref{f:3} lies between the two limits, and is larger for the low-$Z$ case, which attains a value close to 2.5.

To test the extent to which the \napo reaction accounts for the metallicity dependence of $M_\mathrm{Ca}/M_\mathrm{S}$ and $M_\mathrm{Ar}/M_\mathrm{S}$ in type Ia supernova models, I ran one-dimensional SNIa models with a range of progenitor metallicities and the \napo reaction switched off, that is,
\begin{equation}
R_{\mnapo}=f_0 \rho N_\mathrm{Av}\left<\sigma v\right>Y(^{16}\mathrm{O})Y(\mathrm{p})\,,
\end{equation}
\noindent with $f_0 = 0$. The code used is the same as in \citet{2019bra}, where it is described in detail. The thermonuclear reaction rates used in the simulations are those recommended by the JINA REACLIB compilation \citep[][hereafter REACLIB]{2010cyb}. Detailed balance is assumed to hold for forward and reverse reactions, that is, the factor $f_0$ is applied as well to the \isotope{13}{N}$(\alpha,\mathrm{p})$\isotope{16}{O} rate.

\begin{figure}
\resizebox{\hsize}{!}{\includegraphics{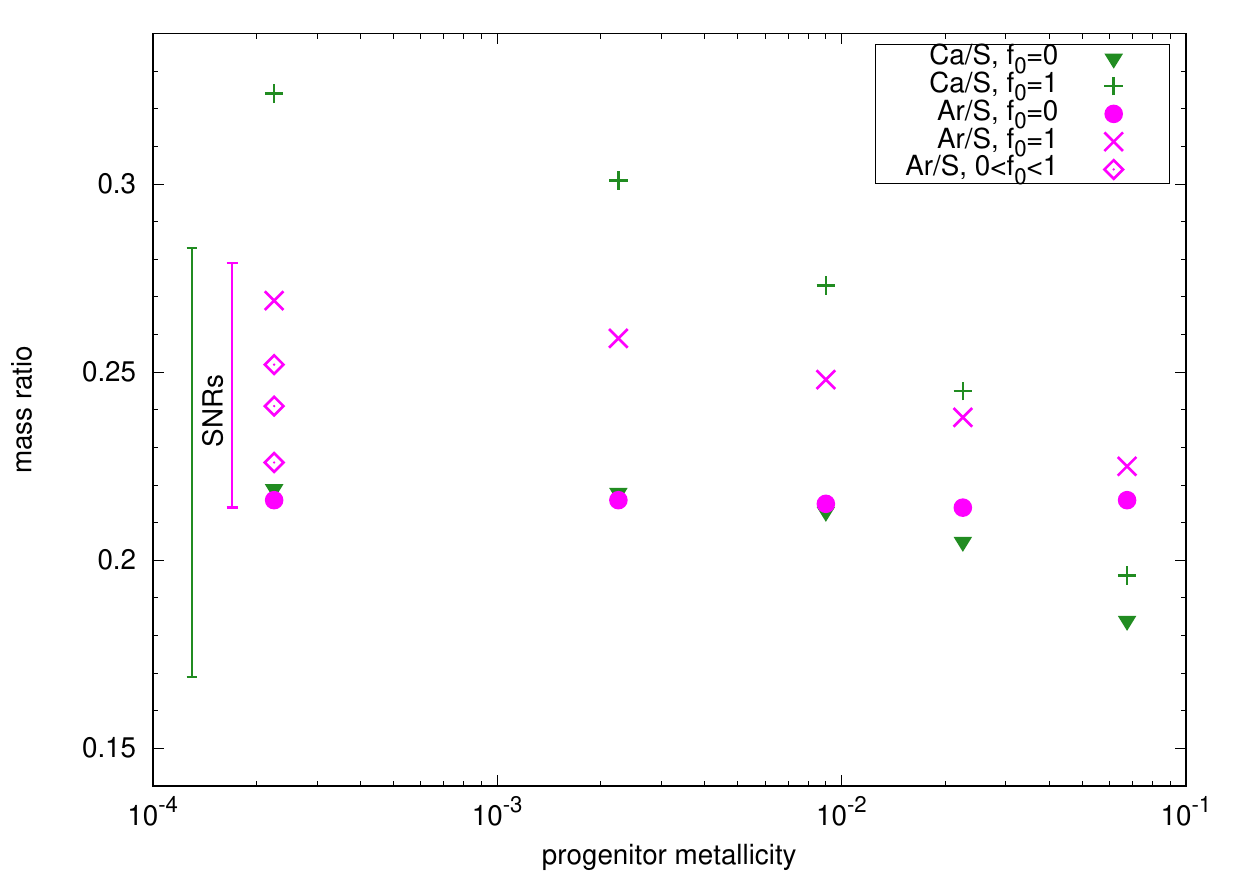}}
\caption{Theoretical mass ratios of calcium to sulfur (in green) and argon to sulfur (in magenta) from models of detonation of a 1.06~\msunb WD, as function of the progenitor metallicity, with either the standard \napo reaction rate ($f_0=1$) or the rate switched off ($f_0=0$). The argon-to-sulfur mass ratio for $Z=2.25\times10^{-4}$ is shown as well for several other values of $f_0$, from bottom to top: $f_0=0.1$, 0.3, and 0.5. The vertical bars on the left of the figure show the range of observational mass ratios derived from measurements of X-ray emission of SNRs \citep{2017mar}. 
}
\label{f:4}
\end{figure}

The results are shown in Fig.~\ref{f:4}, together with the results obtained with the standard rates ($f_0=1$) and the observational constraints derived from the emission lines of SNRs as measured with {\it Suzaku} \citep{2017mar}. Switching off the \napo reaction rate makes $M_\mathrm{Ca}/M_\mathrm{S}$ and $M_\mathrm{Ar}/M_\mathrm{S}$ almost insensitive to the WD progenitor metallicity, at all $Z$ for the mass ratio of argon to sulfur, and at solar and sub-solar $Z$ for the mass ratio of calcium to sulfur. The figure highlights the fact that without the \napo reaction rate these mass ratios cannot cover the full range of measured values at SNRs for any metallicity. The same conclusion holds for different scalings of the \cofus reaction rate, and for models of delayed detonation of Chandrasekhar-mass WDs. 

\section{Limits to the rate of \napo deduced from supernova remnants}\label{s:limits}

At present, the uncertainty on the \napo reaction rate remains unconstrained. Its rate can be found in the STARLIB \citep{2013sal} and REACLIB \citep{2010cyb} compilations. In STARLIB, this rate was computed using Hauser-Feshbach theory and assigned a conventional (recommended) uncertainty of a factor ten, because of the lack of enough experimental information. The REACLIB rate is a fit to the rate of \napo in \citet{1988cau}, and is the rate used in the SNIa models reported here. In the temperature range of interest for explosive oxygen burning, $T\simeq(3.5 - 5)\times10^9$~K, the STARLIB rate is larger than the REACLIB rate by a factor that varies between 1.5 and 2.5.

\begin{figure}
\resizebox{\hsize}{!}{\includegraphics{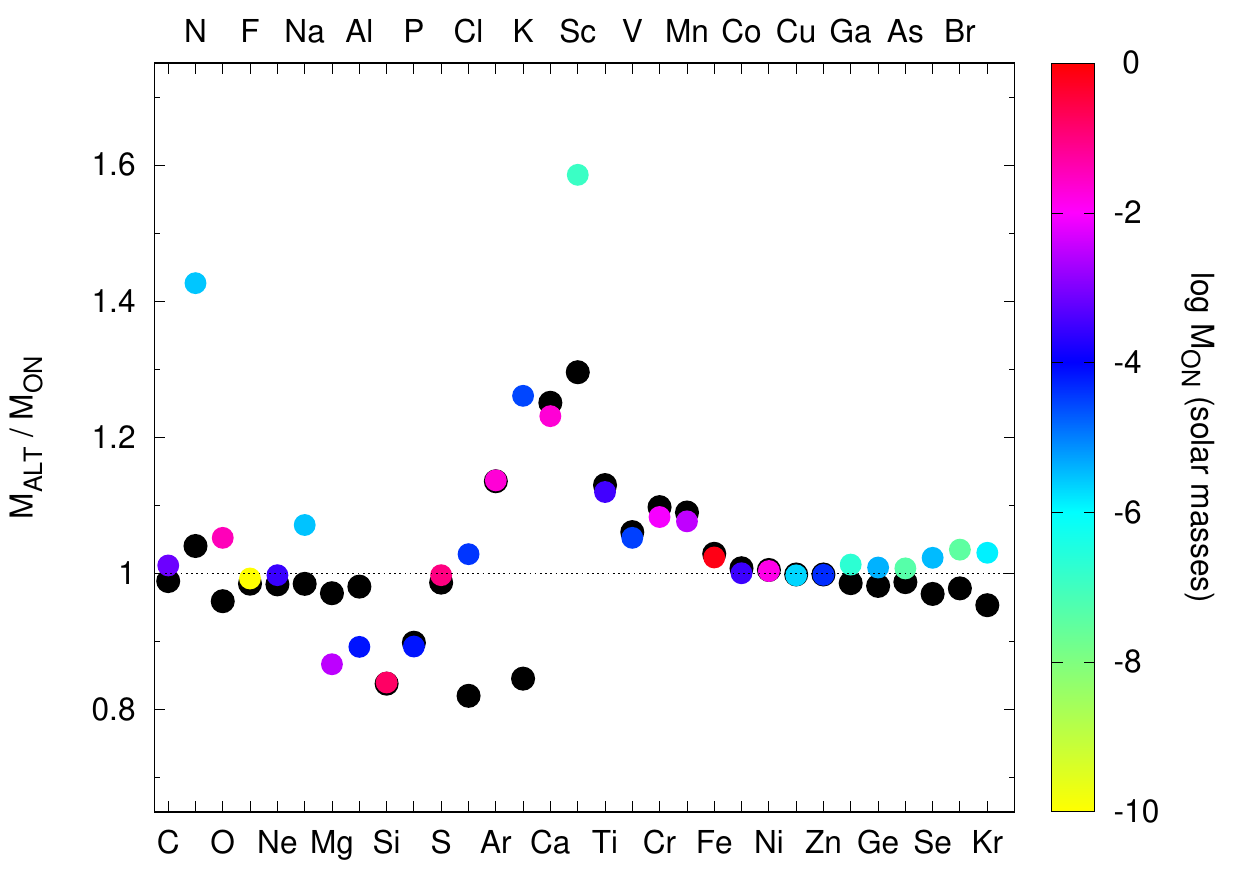}}
\caption{ 
Relative change in the elemental yields obtained in the detonation of a 1.06~\msunb WD, with progenitor metallicity $Z=0.009$, derived from using alternative reaction rates (M$_\mathrm{ALT}$) instead of the standard ones M$_\mathrm{ON}$. The solid coloured 
circles belong to the model with the \cofus reaction rate scaled down by a factor 0.1, and the colour is assigned as a function of the 
yield of each element in model M$_\mathrm{ON}$. The black circles belong to the model with the standard \cofus reaction rate and the \napo reaction rate enhanced by a factor $f_0=7$. 
}
\label{f:5}
\end{figure}

An enhanced \napo reaction rate may also increase the calcium-to-sulfur mass ratio, and becomes an alternative to the scaling down of the \cofus reaction rate by a factor 0.1 suggested by \citet{2017mar} in order to match the range of $M_\mathrm{Ca}/M_\mathrm{S}$ and $M_\mathrm{Ar}/M_\mathrm{S}$ in SNRs. This is because the \napo reaction and its inverse are not in statistical equilibrium at the temperatures reached during explosive oxygen burning, unlike most of the reactions linking intermediate-mass nuclei from silicon to calcium. 

Figure~\ref{f:5} shows the relative change in the elemental yields of the SNIa model consisting in the detonation of a 1.06~\msunb WD with $Z=0.009$, when either the \cofus rate is scaled down by a factor ten or when the \napo rate is scaled up by a factor seven (both models named as M$_\mathrm{ALT}$ in the plot), compared to the same model with all the rates at their standard values (identified in the plot as M$_\mathrm{ON}$). The graph shows that the elements synthesized in significant quantities in the SNIa model, iron-group elements plus intermediate-mass $\alpha$-nuclei, are made in equal proportions in the two M$_\mathrm{ALT}$ models. The same result is obtained for different parameters of the SNIa model; for example, the WD progenitor metallicity. In practice, it is possible to obtain the same proportions of the most abundant elements with intermediate modifications of both the \cofus and the \napo reaction rates. 

\begin{figure}
\resizebox{\hsize}{!}{\includegraphics{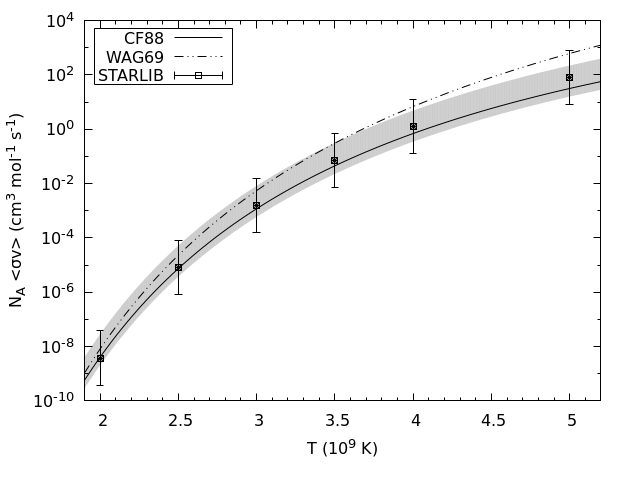}}
\caption{Rate of the reaction $^{16}$O$($p,$\alpha)^{13}$N, as a function of temperature in units of $10^9$~K, from different compilations: \citet[][CF88]{1988cau}, \citet[][WAG69]{1969wag}, and \citet[][STARLIB]{2013sal}. The uncertainty listed in the STARLIB compilation is plotted as a vertical error bar. The shaded band shows the uncertainty on the rate as determined using measured calcium-to-sulfur and argon-to-sulfur mass ratios in SNRs (see text for details).
}
\label{f:6}
\end{figure}

The strong suppression of the metallicity dependence of $M_\mathrm{Ca}/M_\mathrm{S}$ and $M_\mathrm{Ar}/M_\mathrm{S}$ when the \napo reaction is switched off suggests that there should be a minimum for its rate, below which the SNR measurements could not be reproduced. I ran the same model of the detonation of a 1.06~\msunb WD with $Z=2.25\times10^{-4}$ (the effect is most evident at low metallicities) and the \napo reaction rate scaled by different factors, $f_0=0.1$, 0.3, and 0.5. The results are shown in Fig.~\ref{f:5}. The mass ratios belonging to $f_0=0.1$ and 0.3 fall short of covering the observational data, while the results for $f_0=5$ are acceptable, in the sense that the resulting $M_\mathrm{Ar}/M_\mathrm{S}$ is within $3\sigma$ of the upper limit of the corresponding observational range \citep[the $1\sigma$ uncertainty of the upper limit of the argon-to-sulfur mass ratio is 0.01; see][]{2017mar}. Therefore, I chose the last factor, $f_0=0.5$, to establish a lower limit to the \napo reaction rate.

The restrictions to the \napo reaction rate derived from the measurements of $M_\mathrm{Ca}/M_\mathrm{S}$ and $M_\mathrm{Ar}/M_\mathrm{S}$ in SNRs are displayed in Fig.~\ref{f:6}, together with the rates from STARLIB, REACLIB, and from \citet{1969wag}, the one used by \citet{1973woo}. The SNR observational data lead to tighter rate constraints than the uncertainty listed in STARLIB, although the rates provided by this compilation lie within the observationally based rate uncertainty (shaded band in Fig.~\ref{f:6}). On the other hand, the rates from \citet{1969wag} are too large at high temperatures. It is to be expected that future X-ray observatories will be able to provide more stringent constraints on this rate through more accurate data concerning the strength of the emission lines of intermediate-mass elements in SNRs.

\section{Conclusions}\label{s:conclusions}

The so-called $\alpha$-rich explosive oxygen burning during type Ia supernova explosions enhances the production of calcium with respect that of sulfur. From previous studies, it is known that there are two effects to account for in relation to $\alpha$-rich oxygen burning. First, the strength of the enhancement of the yield of calcium at all metallicities, and second, the metallicity dependence of the mass ratio of calcium to sulfur, $M_\mathrm{Ca}/M_\mathrm{S}$, in the ejecta.

Here, it is demonstrated that a single reaction, \napo (followed by \isotope{13}{N}$+\gamma\rightarrow\mathrm{p}+$\isotope{12}{C}), is responsible for the metallicity dependence of $M_\mathrm{Ca}/M_\mathrm{S}$ in the ejecta of type Ia supernovae. This reaction chain boosts $\alpha$-rich oxygen burning when proton abundance is large, increasing the synthesis of argon and calcium with respect to sulfur and silicon. For high-metallicity progenitors, the presence of free neutrons leads to a drop in the proton abundance and the above chain is not efficient. Through one-dimensional modeling of supernova explosions, it is shown that switching off the \napo rate makes the nucleosynthesis insensitive to the metallicity of the supernova progenitor. 

Although the rate of \napo can be found in astrophysical reaction rate libraries, its uncertainty is unconstrained. Assuming that all reaction rates other than \napo retain their standard values, an increase by a factor of approximately seven  of the \napo rate at temperatures in the order $3-4\times10^9$~K is enough to explain the whole range of calcium-to-sulfur mass ratios measured in Milky Way and LMC supernova remnants. These same measurements provide a lower limit to the \napo rate in the mentioned temperature range, on the order of a factor 0.5 with respect to the rate reported by Caughlan \& Fowler in 1988. Future measurements of the \napo rate at the energies of the Gamow-peak for temperatures in the range $3-4\times10^9$~K are encouraged, as they would help to determine the precise role of this reaction in the synthesis of calcium in type Ia supernovae.

\begin{acknowledgements}
This work has benefited from discussions about explosive oxygen burning with Frank Timmes, Broxton Miles, Dean Townsley, Carles Badenes, and H\'ector Mart\'\i nez-Rodr\'\i guez.
Support by the MINECO-FEDER grant AYA2015-63588-P is acknowledged.
\end{acknowledgements}

\bibliographystyle{aa}
\bibliography{../../ebg}

\begin{thebibliography}{28}
\expandafter\ifx\csname natexlab\endcsname\relax\def\natexlab#1{#1}\fi

\bibitem[{{Ashall} {et~al.}(2016){Ashall}, {Mazzali}, {Pian}, \&
  {James}}]{2016ash}
{Ashall}, C., {Mazzali}, P.~A., {Pian}, E., \& {James}, P.~A. 2016, \mnras,
  463, 1891

\bibitem[{{Badenes} {et~al.}(2008){Badenes}, {Bravo}, \& {Hughes}}]{2008bad}
{Badenes}, C., {Bravo}, E., \& {Hughes}, J.~P. 2008, \apjl, 680, L33

\bibitem[{{Bravo} {et~al.}(2019){Bravo}, {Badenes}, \&
  {Mart{\'{\i}}nez-Rodr{\'{\i}}guez}}]{2019bra}
{Bravo}, E., {Badenes}, C., \& {Mart{\'{\i}}nez-Rodr{\'{\i}}guez}, H. 2019,
  \mnras, 482, 4346

\bibitem[{{Bravo} \& {Mart{\'i}nez-Pinedo}(2012)}]{2012bra}
{Bravo}, E. \& {Mart{\'i}nez-Pinedo}, G. 2012, \prc, 85, 055805

\bibitem[{{Caughlan} \& {Fowler}(1988)}]{1988cau}
{Caughlan}, G.~R. \& {Fowler}, W.~A. 1988, Atomic Data and Nuclear Data Tables,
  40, 283

\bibitem[{{Chamulak} {et~al.}(2008){Chamulak}, {Brown}, {Timmes}, \&
  {Dupczak}}]{2008chm}
{Chamulak}, D.~A., {Brown}, E.~F., {Timmes}, F.~X., \& {Dupczak}, K. 2008,
  \apj, 677, 160

\bibitem[{{Cyburt} {et~al.}(2010){Cyburt}, {Amthor}, {Ferguson}, {Meisel},
  {Smith}, {Warren}, {Heger}, {Hoffman}, {Rauscher}, {Sakharuk}, {Schatz},
  {Thielemann}, \& {Wiescher}}]{2010cyb}
{Cyburt}, R.~H., {Amthor}, A.~M., {Ferguson}, R., {et~al.} 2010, \apjs, 189,
  240

\bibitem[{{De} {et~al.}(2014){De}, {Timmes}, {Brown}, {Calder}, {Townsley},
  {Athanassiadou}, {Chamulak}, {Hawley}, \& {Jack}}]{2014de}
{De}, S., {Timmes}, F.~X., {Brown}, E.~F., {et~al.} 2014, \apj, 787, 149

\bibitem[{{Fesen} {et~al.}(1988){Fesen}, {Wu}, {Leventhal}, \&
  {Hamilton}}]{1988fes}
{Fesen}, R.~A., {Wu}, C.-C., {Leventhal}, M., \& {Hamilton}, A.~J.~S. 1988,
  \apj, 327, 164

\bibitem[{{Hamilton} \& {Fesen}(1988)}]{1988ham}
{Hamilton}, A.~J.~S. \& {Fesen}, R.~A. 1988, \apj, 327, 178

\bibitem[{{Hughes} {et~al.}(1995){Hughes}, {Hayashi}, {Helfand}, {Hwang},
  {Itoh}, {Kirshner}, {Koyama}, {Markert}, {Tsunemi}, \& {Woo}}]{1995hug}
{Hughes}, J.~P., {Hayashi}, I., {Helfand}, D., {et~al.} 1995, \apjl, 444, L81

\bibitem[{{Lodders}(2003)}]{2003lod}
{Lodders}, K. 2003, \apj, 591, 1220

\bibitem[{{Mart{\'{\i}}nez-Rodr{\'{\i}}guez}
  {et~al.}(2017){Mart{\'{\i}}nez-Rodr{\'{\i}}guez}, {Badenes}, {Yamaguchi},
  {Bravo}, {Timmes}, {Miles}, {Townsley}, {Piro}, {Mori}, {Andrews}, \&
  {Park}}]{2017mar}
{Mart{\'{\i}}nez-Rodr{\'{\i}}guez}, H., {Badenes}, C., {Yamaguchi}, H.,
  {et~al.} 2017, \apj, 843, 35

\bibitem[{{Mart{\'{\i}}nez-Rodr{\'{\i}}guez}
  {et~al.}(2016){Mart{\'{\i}}nez-Rodr{\'{\i}}guez}, {Piro}, {Schwab}, \&
  {Badenes}}]{2016mar}
{Mart{\'{\i}}nez-Rodr{\'{\i}}guez}, H., {Piro}, A.~L., {Schwab}, J., \&
  {Badenes}, C. 2016, \apj, 825, 57

\bibitem[{{Mazzali} {et~al.}(2008){Mazzali}, {Sauer}, {Pastorello}, {Benetti},
  \& {Hillebrandt}}]{2008maz}
{Mazzali}, P.~A., {Sauer}, D.~N., {Pastorello}, A., {Benetti}, S., \&
  {Hillebrandt}, W. 2008, \mnras, 386, 1897

\bibitem[{{Miles} {et~al.}(2016){Miles}, {van Rossum}, {Townsley}, {Timmes},
  {Jackson}, {Calder}, \& {Brown}}]{2016mil}
{Miles}, B.~J., {van Rossum}, D.~R., {Townsley}, D.~M., {et~al.} 2016, \apj,
  824, 59

\bibitem[{{Piersanti} {et~al.}(2017){Piersanti}, {Bravo}, {Cristallo},
  {Dom{\'i}nguez}, {Straniero}, {Tornamb{\'e}}, \&
  {Mart{\'i}nez-Pinedo}}]{2017pie}
{Piersanti}, L., {Bravo}, E., {Cristallo}, S., {et~al.} 2017, \apjl, 836, L9

\bibitem[{{Piro} \& {Bildsten}(2008)}]{2008pir}
{Piro}, A.~L. \& {Bildsten}, L. 2008, \apj, 673, 1009

\bibitem[{{Sallaska} {et~al.}(2013){Sallaska}, {Iliadis}, {Champange},
  {Goriely}, {Starrfield}, \& {Timmes}}]{2013sal}
{Sallaska}, A.~L., {Iliadis}, C., {Champange}, A.~E., {et~al.} 2013, \apjs,
  207, 18

\bibitem[{{Sasdelli} {et~al.}(2014){Sasdelli}, {Mazzali}, {Pian}, {Nomoto},
  {Hachinger}, {Cappellaro}, \& {Benetti}}]{2014sas}
{Sasdelli}, M., {Mazzali}, P.~A., {Pian}, E., {et~al.} 2014, \mnras, 445, 711

\bibitem[{{Stehle} {et~al.}(2005){Stehle}, {Mazzali}, \&
  {Hillebrandt}}]{2005sth}
{Stehle}, M., {Mazzali}, P.~A., \& {Hillebrandt}, W. 2005, Nuclear Physics A,
  758, 470

\bibitem[{{Tanaka} {et~al.}(2011){Tanaka}, {Mazzali}, {Stanishev}, {Maurer},
  {Kerzendorf}, \& {Nomoto}}]{2011tan}
{Tanaka}, M., {Mazzali}, P.~A., {Stanishev}, V., {et~al.} 2011, \mnras, 410,
  1725

\bibitem[{{Truran} \& {Arnett}(1970)}]{1970tru}
{Truran}, J.~W. \& {Arnett}, W.~D. 1970, \apj, 160, 181

\bibitem[{{Vancura} {et~al.}(1995){Vancura}, {Gorenstein}, \&
  {Hughes}}]{1995van}
{Vancura}, O., {Gorenstein}, P., \& {Hughes}, J.~P. 1995, \apj, 441, 680

\bibitem[{{Wagoner}(1969)}]{1969wag}
{Wagoner}, R.~V. 1969, \apjs, 18, 247

\bibitem[{{Woosley} {et~al.}(1973){Woosley}, {Arnett}, \& {Clayton}}]{1973woo}
{Woosley}, S.~E., {Arnett}, W.~D., \& {Clayton}, D.~D. 1973, \apjs, 26, 231

\bibitem[{{Yamaguchi} {et~al.}(2015){Yamaguchi}, {Badenes}, {Foster}, {Bravo},
  {Williams}, {Maeda}, {Nobukawa}, {Eriksen}, {Brickhouse}, {Petre}, \&
  {Koyama}}]{2015yam}
{Yamaguchi}, H., {Badenes}, C., {Foster}, A.~R., {et~al.} 2015, \apjl, 801, L31

\bibitem[{{Yamaguchi} {et~al.}(2014){Yamaguchi}, {Badenes}, {Petre}, {Nakano},
  {Castro}, {Enoto}, {Hiraga}, {Hughes}, {Maeda}, {Nobukawa}, {Safi-Harb},
  {Slane}, {Smith}, \& {Uchida}}]{2014yam}
{Yamaguchi}, H., {Badenes}, C., {Petre}, R., {et~al.} 2014, \apjl, 785, L27

\end{thebibliography}

\end{document}